\documentclass[conference, one column]{IEEEtran}
\pdfoutput=1 
\usepackage{amsmath}
\usepackage{amsfonts}
\usepackage{amssymb}
\usepackage{multirow}
\usepackage{graphicx}

\begin{document}
\title{Secure Data Storage Structure and Privacy-Preserving Mobile Search Scheme for Public Safety Networks}
\author{\IEEEauthorblockN{Hamidreza Ghafghazi\IEEEauthorrefmark{1}, Amr ElMougy\IEEEauthorrefmark{2}, Hussein T. Mouftah\IEEEauthorrefmark{1}, Carlisle Adams\IEEEauthorrefmark{1}}\\
\IEEEauthorblockA{Department of Electrical Engineering and Computer Science, University of Ottawa\IEEEauthorrefmark{1}\\German University in Cairo\IEEEauthorrefmark{2}\\Emails: \{hamidreza.ghafghazi, mouftah, cadams\}@uottawa.ca\IEEEauthorrefmark{1}, amr.elmougy@guc.edu.eg\IEEEauthorrefmark{2}
}}
\maketitle

\begin{abstract}
In a Public Safety (PS) situation, agents may require critical and personally identifiable information. 
Therefore, not only does context and location-aware information need to be available, but also the privacy of such information should be preserved. Existing solutions do not address such a problem in a PS environment. This paper proposes a framework in which anonymized Personal Information (PI) is accessible to authorized public safety agents under a PS circumstance. 
In particular, we propose a secure data storage structure along with privacy-preserving mobile search framework, suitable for Public Safety Networks (PSNs). As a result, availability and privacy of PI are achieved simultaneously. However, the design of such a framework encounters substantial challenges, including scalability, reliability of the data, computation and communication and storage efficiency, etc. We leverage Secure Indexing (SI) methods and modify Bloom Filters (BFs) to create a secure data storage structure to store encrypted meta-data. As a result, our construction enables secure and privacy-preserving multi-keyword search capability. In addition, our system scales very well, maintains availability of data, imposes minimum delay, and has affordable storage overhead. We provide extensive security analysis, simulation studies, and performance comparison with the state-of-the-art solutions to demonstrate the efficiency and effectiveness of the proposed approach. To the best of our knowledge, this work is the first to address such issues in the context of PSNs.

\end{abstract}
\section{Introduction}
 
Suppose a large building (e.g., with 10 floors) is on fire and many people are trapped inside, or a region with a population of 1000 civilians and 300 houses faces a natural incident like an earthquake or flood. Under such circumstances, Public Safety Agents (PSAs) seek all possible information to achieve Situational Awareness (SA) \cite{Ghafghazi:2014:CTP:2642687.2642693}. In general, they would like to find answers to the following questions: How many people are in danger? Is there a way to identify those individuals? What is the closest location of the endangered individuals? What were the health condition of people before the incident? What are their health condition at the moment? Where can we find the health records of endangered individuals? Is there anybody except PSAs who are close to the situation and who have certain capabilities that can be used to help people, like engineers, physicians, nurses, etc? Many previous incidents have shown that people care about each other in hard times and, in fact, volunteer to help others in need. In this case, another question would be, how is it possible to reach out to those available volunteers in critical situations and ask them for their help? The answers to the aforementioned questions can help PSAs to prioritize their rescue missions, which can result in saving many lives. 

However, there are substantial challenges towards designing a system which addresses such requirements. First, Public Safety (PS) situations are highly dynamic. This mandates strict requirements including high scalability, data availability, and very low response time. This is because in such cases a large number of people may be affected, among which many may require immediate care. Second, the sought information is vastly distributed which makes data retrieval process even more complex. For example, Physical Health Records (PHRs) are stored in proprietary hospitals, or in various cloud servers like Amazon, Google, and Microsoft to name a few. In fact, before retrieving any Personal Information (PI) like PHRs, it is necessary to identify endangered individuals and the servers to which they have outsourced their information. Without proper identification, no information can be retrieved. Third, the sought information is considered Personally identifiable information which raises privacy issues.\cite{Ghafghazi:2014:CTP:2642687.2642693}

\textbf{Our Contributions}: 
in this work\footnote{This work has been accepted to be presented in Wireless Communications and Networking Conference (WCNC), 2016 IEEE.}, we propose a framework to answer the aforementioned questions raised in the introduction and to tackle the issues highlighted in the related work. Our framework not only provides a sufficient level of SA for PSAs, which results in saving human lives, but also it addresses the pre-requisite step for data retrieval which is privacy-preserving user and server identification. In this regard, we propose a secure data storage structure to store "meta" PI. To provide high data availability, in addition to the cloud storage model, we utilize an opportunistic storage 
to store our data structure in mobile clouds. We propose a privacy-preserving search algorithm to facilitate multi-keyword search for AND/OR queries. The search algorithm imposes minimum delay which is desirable for PS situations. We provide extensive security analysis, simulation studies, and performance comparison with the state-of-the-art solutions to demonstrate the efficiency and effectiveness of the proposed approach. To the best of our knowledge, this work is the first to address such issues in the context of PSNs. Table \ref{Abb} summarizes some abbreviations used in this work.

The remaining sections are as follows. System model, threat model, and assumptions are presented in Section 2. Section 3 elaborates our scheme construction. Security analysis and performance evaluation are discussed in Sections 4 and 5 respectively. Section 6 summarizes the paper. 

\begin{table}[h]
\centering
		\caption{List of Abbreviations }
\begin{tabular}{cc|cc}

\hline Abbreviation & Description & Abbreviation & Description \\
\hline
\hline PS & Public Safety &PSN & Public Safety Network \\ 
 
\hline PSA & Public Safety Agent& SA & Situational Awareness  \\ 
\hline PHR& Physical Health Record &PI & Personal Information  \\ 
\hline MI & Meta Information & BF & Bloom Filter \\
\hline  DO & Data Owner & HSP &Health Service Provider \\ 
\hline $A_C$& Centralized Availability &$A_D$ &Decentralized Availability \\ 
\hline  DA &Direct Authorization &  IA& Indirect Authorization \\ 
\hline KGA & Key Generation Authority & TA & Trusted Authority\\ 
\hline CCS & Central Cloud Server & CS & Cloud Servers\\
\hline MC &Mobile cloud & OE & Obfuscating Element\\
\hline SBF & Storage bloom filter &CBF & Counting bloom filter  \\ 
\hline OBF & Obfuscating bloom filter & RBF & Removal bloom filter\\ 
\hline ABE & Attribute-based Encryption &PDA & Personal Digital Assistant\\ 

\hline 
\end{tabular} 

\label{Abb}
\end{table}

\section{Related Work}
Surveying the literature, the works proposed for centralized and mobile healthcare and emergency handling, and search over encrypted data, are the most germane to ours. In such research fields, data availability is achieved in two ways; Centralized Availability ($A_C$) and Decentralized Availability ($A_D$). In $A_C$, Data Owners (DOs) outsource the encrypted information to one/many cloud server(s) to which PSAs should send information retrieval requests, while in $A_D$, in an emergency, DOs broadcast their encrypted PI using smart phones or Personal Digital Assistants (PDAs) to the users in their local proximity or to Health Service Providers (HSPs) to ask for help. The PDA monitors and collects health information using the sensors attached to the patient's body. To achieve data privacy, in addition to data confidentiality, Direct Authorization (DA) or Indirect Authorization (IA) algorithms are utilized. DA methods are usually used in private domains which are comprised of family, personal physician, friends, and neighbours, while IA is applied in public domains that include researchers, healthcare personnel, other doctors, and so forth \cite{li2011authorized}. 

Although DA and IA highly affect data availability, the details of the authorization process lie outside the scope of this paper. Therefore, we merely describe the high level overview of those processes. 

In $A_C$, to achieve DA, any user who is interested in the patient's information should directly contact her even in an emergency and ask for access authorization. For example, in \cite{samanthula2015secure, sun2011hcpp}, upon a request from a user, the DO sends decryption keys only if the user passes the authorization check phase. Such an approach is impractical in PS situations for two main reasons; firstly, they do not scale well; and secondly, DOs may be unconscious or may not even be reachable to grant access to the users.

On the other hand, IA has been used for the public domain in which DOs either delegate access authorization to a cloud server or enforce access policies into the ciphertext upon outsourcing data. In this case, a user seeking some information, without contacting DOs, sends a request to a server and retrieves the information all at once. Thus, this approach scales well which makes it more suitable for PS situations. 

Tong, et al. \cite{6683024} proposes that DOs delegate the access authorization to a private cloud. This scheme enhances Searchable Symmetric Encryption (SSE) using pseudo-random number generators to avoid linkability of file identities. SSE uses linked lists in which file identities containing similar keywords are linked together in a secure way. The algorithm imposes minimum search delay since it does not need to search over the entire database to find the result. However, its efficiency drops in dynamic situations in which files are added/removed to/from the system frequently. In addition, the scheme is not able to perform multi-keyword search and the private cloud learns the keywords for which a user would like to search the database. 

The work in \cite{liu2012secure} uses the Public-key Encryption with Keyword Search (PEKS) algorithm to preserve keyword privacy. With PEKS, a trapdoor is computed for a keyword and upon search, it is compared against the entire database to find the results.  
However, the scheme is not efficient, firstly, because to retrieve proper information the entire database should be searched, and secondly, it is computationally expensive as PEKS employs pairing-based cryptography (PBC). To tackle the latter, \cite{dong2011shared, liu2012secure} proposed to outsource the heavy computations of PBC to a proxy server. The approach converts a ciphertext in such a way that the decryption process is more lightweight at the user side. Despite the preceding improvement, in a PS environment, the number of data outsourcing requests may be quite large because of the large amount of information. This causes the delay to be increased. Furthermore, in such situations, the network infrastructure might be down which may result in lack of access to the proxy servers. Therefore, the applicability of such techniques is questionable in this context.

To achieve IA, a DO can also enforce access authorization into the ciphertext using functional encryption (for example, Attribute-based Encryption (ABE) or Predicate Encryption (PE)). ABE enables a DO-centric authorization model. In \cite{barua2011peace}, DOs send data to an HSP and delegate access authorization to that entity. Then, the HSP first classifies the data using the attribute set chosen by the DO and then uses ABE to enforce the DO's access policy for the users. The works \cite{zhou2014psmpa,li2013scalable} use ABE and suggest to form an emergency version of encrypted data in which the owner only uses the "emergency" attribute to produce an emergency ciphertext. Then, the DO delegates emergency keys to a trusted authority. In an emergency, healthcare personnel can retrieve the emergency key to decrypt data. The authors in \cite{li2011authorized} propose authorized multi-keyword search using predicate encryption. The delay corresponding to the search process is proportional to the size of the database and it involves pairing computations. Similarly, the preceding schemes are all based on PBC which are suitable for delay tolerant situations; thus, in large settings of a PS environment with delay constraints, those schemes lose their functionality.

However, the $A_C$-related schemes cannot be used directly for PS situations. This is because it is required that endangered individuals, and the servers to which they have uploaded their information, be identified first. Only then are such algorithms able to retrieve information. Without this prior step, one should send a query to all of the target servers. In addition, the queries are going to be very complex since they need to target only the set of PI which belongs to endangered individuals. Note that, although the number of endangered individuals is large compared to a normal situation, their corresponding information encompasses a limited portion of all the data that is stored in such servers. However, a query without such boundaries would result in information about both endangered and safe individuals. This also increases response time. 

In $A_D$, in an emergency, DOs or their PDAs disseminate encrypted PI either to the users within their local proximity or to HSPs. DA is achieved by DOs checking users' legitimacy before data dissemination. If the users passed authorization checks, they would receive encrypted data and the decryption key. However, to achieve IA, DOs enforce access policies into the ciphertext as mentioned above. For the former, the authors in \cite{lu2013spoc} propose to opportunistically use authorized users to outsource health data processing in emergency situations. The authors propose a two-phased access control in which the first phase identifies medical users and the second phase uses a novel scalar product computation algorithm to ensure users' authorization. This approach involves three rounds of communications and involves pairing computation in the first phase of the check process. For the latter, in \cite{liang2011pec, liang2012healthshare} access control is encoded into the ciphertext using ABE and PBC. The authors in \cite{liang2012healthshare} propose direct and indirect transmission modes. The latter delegates data transmission to a more powerful user. The authors in \cite{zhang2014phda} consider priority for different types of health data and perform priority-based data aggregation and transmission. The scheme uses PBC for authorization checks and the paillier cryptosystem for privacy preserving data aggregation. 
The common problem with all of the above schemes is the imposed delay as the result of several rounds of communications \cite{lu2013spoc} or pairing computations. In addition, the schemes are only applicable for individuals who have sensors attached to their bodies for health monitoring and need constant care. In PS situations where the number of individuals involved might be high, they might even be unreachable for some time intervals, and their PDAs or smart phones may be damaged, these methods lack proper functionality. Table \ref{TableI} summarizes the protocol comparison.

\begin{table}[h]
\centering
\caption{Protocol comparison}
\begin{tabular}{ |c|c|c|c| }
\hline
\multirow{1}{*}{Data} & \multirow{1}{*}{Authorization} & \multirow{2}{*}{Scheme} & \multirow{2}{*}{Disadvantages in PS situations}\\ 
\multirow{1}{*}{Availability} & Model &  & \\ \hline

\multirow{8}{*}{$A_C$} & \multirow{2}{*}{DA} & [3] &  \multirow{1}{*}{non-scalable, server and user }\\\cline{3-3}
 &  & [4] & identification, unreachable individuals\\\cline{2-2}\cline{3-3}\cline{4-4}
 & \multirow{6}{*}{IA} & [5] &  \\\cline{3-3}
           &  & [2] & Search delay proportional \\ \cline{3-3}
  &  & [6] & to database size [2,6], User and  \\\cline{3-3}
 &  & [7] & server identification,  \\ \cline{3-3}
  &  & [8] & high computation delay,\\ \cline{3-3}
      &  & [9] & unsearchable data retrieval [7-10]\\ \cline{3-3}
         &  & [10] &\\ \cline{3-3}
 \hline
\multirow{6}{*}{$A_D$} &  \multirow{2}{*}{DA} &  \multirow{2}{*}{[11]} & \multirow{1}{*}{Delay due to three rounds of communication} \\\cline{4-4}
  & & &  Limited to unhealthy individuals, \\\cline{2-2}\cline{3-3}
 & \multirow{3}{*}{IA} & [12]& data unavailability due to device \\\cline{3-3}
 &  & [13]& damage, high computation \\\cline{3-3}
  &  & [14]& delay\\\cline{3-3}\hline
\end{tabular}
\label{TableI}
\end{table}

\section{System model and Threat model}
It is assumed that a city is divided into several distinct zones, each having a unique pseudo-identity $PS_a$. Each zone may have its own cloud server. As an alternative, we can have one central cloud server. Then, we can dedicate a separate data storage structure to each zone both of which share the same pseudo-identity. The area of a zone depends on the number of registered individuals expected to reside inside that area at once and the memory size allocated to the storage structure for each zone. Without loss of generality, in this work we assume that one Central Cloud Server (CCS) stores all distinct data structures. The system is comprised of several entities as follows.

\textbf{Key Generation Authority (KGA)}: This entity generates the secret keys of the system.

\textbf{Cloud Servers (CS)}: In addition to CCS, we assume that there are several cloud servers, each managed by different vendor such as Google, Microsoft, Amazon, and so forth. These servers store the complete version of PI for individuals which may be up to 200 pages per record \cite{6560004}. 
\textbf{DO}: This entity is a member of the general public. People upload their information such as PHRs to central and mobile clouds.

\textbf{PSAs}: these are the governmental authorities including policemen, firefighters, and paramedics. 

\textbf{Mobile Cloud (MC)}: we assume that there are individuals who possess powerful smartphones with computing capabilities and spare storage. Individuals in their local proximity will upload their information to these providers either before an incident or during one. 

\textbf{Adversary}: This entity will try to eavesdrop on the communications and send queries to CCS to retrieve information. This work does not focus on DoS.  
\begin{figure}[tp]
  \centering
  \includegraphics[width=\textwidth, height=!]{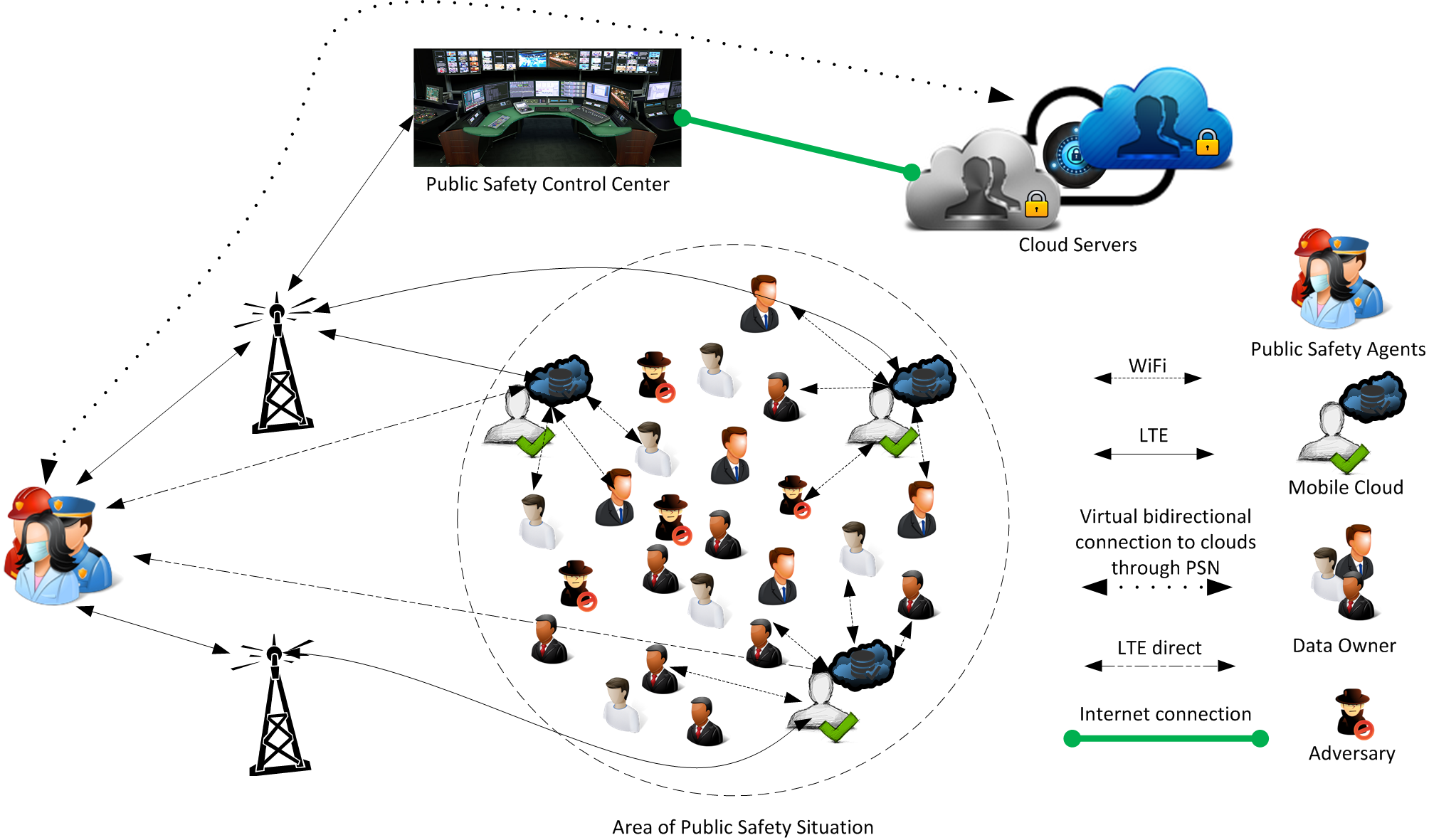}
  \caption{System model}
  \label{system model1}
  \end{figure}

As depicted in Fig \ref{system model1}, there is an area in which a PS situation has occurred.
Note that before an incident happens and during normal conditions, those DOs who have registered to the system will outsource their encrypted Meta Information (MI) using our algorithm to one or many central and mobile cloud provider(s). 
MI is comprised of Health Attribute (HA) keywords set (e.g., $H_u=\{Asthma, A^+, etc.\}$), a pseudonym of the DO $U_{P_{ID}}$, the identity of a CS where a complete version of their PHR is stored (e.g., $CS_x$), the memory index $\sigma _u$ where the PHR is recorded, and other kinds of PI relevant to an emergency $PI_{em}$ (e.g., emergency contact number, civil engineer, electrician, etc.). It has the following format, $MI=\{U_{P_{ID}} || H_u|| CS_x||\sigma _u||PI_{em} \}$. This information is generated by DOs and uploaded to the CCS or a MC. In a PS situation, PSAs can search for distinctive/conjunctive $HAs$ and $PI_{em}$ keywords for a particular affected area, and retrieve such information.

In this work, we assume that KGA is fully trusted, but CSs and MCs are honest but curious. This means that they follow the procedure of the scheme in an honest way, but try to learn as much information as possible. We assume that KGA authenticates DOs and PSAs and only then it transfers secret keys to those entities. However, authentication procedure is out of the scope of this work. We also assume that there exist eavesdropper adversaries who live among the general public and would like to learn as much information as possible. The adversaries may also register to the system and would like to upload their bogus information to tamper with the fast processing of the information and decrease the validity ratio of the information.

\section{Secure Data Storage Structure and Privacy-Preserving Mobile Search Scheme}

\subsection{Storage Bloom Filter}
A Bloom Filter (BF) is a type of data structure that represents a set of $l$ keywords by an array of $m$ bits \cite{broder2004network}. The inputs are as follows:
\begin{enumerate}
\item A set of $l$ keywords: $W=\{w_1,w_2,...,w_l\}$
\item An independent set of hash functions:$\{h_i\}^{r} _{i=1}$ where $h_i: \{0,1\}^* \rightarrow [1,m]$ for $1\leq i\leq r$.
\item A bit array of size $m$ which is set to $0$ initially. 
\end{enumerate}
For each element $w_j\in W$, the bits in $m$ at positions $ h_i(w_j)$ are set to 1 for $ i=1,...,r $. The relationship between the parameters is $m= \frac{lr}{\ln 2}$. To check whether a word $S$ is in the BF or not, we may check $ h_i(S)$ for $ i=1,...,r $; if all of the resultant bits are 1, then $S$ was included in the BF with high probability. Otherwise, even if only one position is $0$, definitely $S$ was not inserted.

We modify a variation of BF called counting BF (CBF) to achieve Storage BF (SBF). CBF is a variation of the standard BF in which each bit of the BF acts as a counter \cite{broder2004network}. Note that in regular BF, any bit can be targeted more than once during the indexing process. But, once it has flipped to 1, its value dose not increment further. Now, to build an SBF, we assume that instead of counters in CBF, we have a set of buffers $B=\{B_1, B_2,...,B_m\}$ to store data. Therefore, to insert a pair $(w_i,\alpha _i)$, $ \forall w_i\in W$ into an SBF, $\alpha _i$ is added to the $B_{h_j(w_i)}$ for $j=1,...,r$ where $B_{h_j(w_i)}$ is the targeted buffer. Then, to check whether $w'_i\in B$ or not, one can check all the sets $B_{h_j(w'_i)}$ and if all the sets are non-empty it returns the value associated with $w'_i$ which is the intersection of all $B_{h_j(w'_i)}$ (i.e., $\bigcap _{j=1} ^r B_{h_j(w'_i)}$). We modify the work in \cite{goh2003secure} and include SBF to the scheme to produce a secure data storage structure and privacy-preserving mobile search algorithm.

\subsection{Construction}
\textbf{Setup($s$):} Given the security parameter $s$, choose a pseudo-random function $f: \{0,1\}^n \times \{0,1\}^s \rightarrow \{0,1\}^s$ and a fixed set of keywords, (e.g., \textit{W =$\{$Asthma, Heart condition, artificial leg, blood type, civil engineer, electrician, etc.$\}$} where $|W|=l$ and $W_i\in \{0,1\}^n$. For each $W_i$, KGA generates a secret key $\varsigma _i \in \{0,1\}^{s}$, $\forall i\in W$. KGA also generates a public-private key pair for PSAs. However, we have extended the preceding and considered different authorization levels for PSAs; this will be presented in a separate work. 

In addition, a set of initial vectors $V=(v_1,v_2,...,v_r)\in \{0,1\}^{nr}$ will be produced. Each user also has a CBF ($CBF_u$), and a standard BF ($BF_u$), both of which will be initialized with 0. On the server side, we utilize SBF to store DOs' data. It is assumed that all buffers of SBF are of the same size. PSAs will receive the key sets with respect to their authorization.  

\textbf{Registration($H_u$):} 
Upon DO registration and based on her set of keywords $W_u=\{H_u\bigcup PI_{em}\}$, KGA will transfer a set of master keys $\varsigma =\{\varsigma_j\}$ for $1\leq j\leq |W_u|$, along with, $V$ to the DO. Then, the DO generates his master keys as in equation \ref{KG} for all $W_j \in W_u$. Note that a master key is in the form of $K_{W_j}=(k_{1,W_j},...,k_{r,W_j})$. It is assumed that $t$ users will sign up to the system. 
\begin{equation}\label{KG}
K_{{W_j}}=(f(v_1, \varsigma _j),f(v_2, \varsigma _j),...,f(v_r,\varsigma _j))\in \{0,1\}^{sr}
\end{equation}

\textbf{BuildIndex$(W_u, K_{u})$:} The input includes the set $W_u$ as keywords and their corresponding master keys. The outputs are $CBF_u$, $BF_u$, and $OBF_b$.

\textit{Step 1}: For every $W_j \in W_u$, compute the following:
\begin{itemize}

\item[(a)] Trapdoor: $T_{W_j}=\{z_1, z_2, ...,z_r\}\in \{0,1\}^{sr}$ is calculated as follows:

$T_{W_j}=\{f(W_j,k_{1,W_j}),..., f(W_j, k_{r,W_j})\}$

\item[(b)] Location vector: if $\gamma\in \{0,1\}^n$ is a specific location inside $PS_a$, then $L_{W_i}=\{y_1,...,y_r\}\in \{0,1\}^{sr}$ is computed as follows:
$L_{W_j}=\{f(\gamma, z_1),...,f(\gamma, z_r)\}$ 
\item[(c)] Insert $y_1,...,y_r$ to both $BF_u$ and $CBF_u$ as follows: 
\begin{equation}\label{m4}
\left \{ \begin{array}{rr}
BF_u: &\quad b_{h_i{(y_i)}}= 1\\
CBF_u: & \quad C_{h_i{(y_i)}}=C_{h_i{(y_i)}}+1
       \end{array} \right.
\end{equation}
Where $i=1,...,r$. Note that $b_{h_i{(y_i)}}$ and $C_{h_i{(y_i)}}$ are bit location and counter location in $BF_u$ and $CBF_u$ respectively. And, $BF_u$ and $CBF_u$ both have the same length. 
\end{itemize}

\textit{Step 2}: We will build an Obfuscating BF ($OBF$) for some extra blinding elements that is used to obfuscate $BF_u$. Suppose $max\{|W_u|\}=q<l $. For a DO, $|W_u| = d\leq q$ for which he builds equation \ref{m4}.  
Then, the DO picks $(q-d)\in \{0,1\}^*$ random values, computes equation \ref{m4}, and only keeps $BF_u$ which we name $OBF$. The DO will update $BF_u$ by the bitwise OR operation of the two (i.e. $BF_u=BF_u \vee OBF$ where $\vee$ represents the bitwise OR operation). Thus, every DO will have the same number of elements to add to the SBF. 

\textit{Step 3}: The output is $\, I_U=(CBF_u, BF_u,OBF)$. Using data compression tools, one can drastically decrease the size of the $BF_u$ to be sent. This is because most of the bits in it are zero for a single user. Thus, the DO calculates a compression function with $BF_u$ as an input, $C(BF_u)$. Then, he/she sends a packet to the CCS/MCs in the form of $\xi ' \{\xi (MI_u)|| C(BF_u)|| PS_a\}$ where $||$ means concatenation.
The server simply decrypts the packet, decompresses $C(BF_u)$, and adds $\xi (MI_u)$ to the corresponding buffers in the $SBF$ using $BF_u$. Note that using the PSAs' public key and a secure public-key encryption scheme, we generate $\xi(MI_u)$. Also, note that it is assumed that a DO and a CCS/MC use SSL to secure communications between one another. 

\textbf{SearchIndex$(W_i, \gamma)$:} The input is the trapdoor for a particular $W_i$ and the specific location for where PSAs seek information. The output is a set of $MI_u$ associated with $W_i$. 
\begin{itemize}
\item [(a)] For a $W_i$, a PSA follows the step 1 of BuildIndex to calculate $T_{W_i}$ and $L_{W_i}$, then send $L_{W_i}=\{f(\gamma, z_1),...,f(\gamma, z_r)\}$ in an encrypted form to the CCS/MC. 

\item [(b)] Then, the CCS/MC finds $\nu=\bigcap_{i=1} ^r B_{h_i (y_i)}$, $ \forall B_{h_i (y_i)}\in SBF$, and send back $\nu$ to the PSA. 
\item [(c)] The PSA will decrypt each element of $\nu$ to find $MI_u$ for all the DOs with the same keyword. If necessary, the PSAs are able to retrieve the complete PI using $CS_x||\sigma _u$.

\end{itemize}
 
\subsection{Add/Remove and Multi-Keyword Search}

Features like addition and removal enable DOs to update the SBF when they move from one location to another. They are supported in our construction using $CBF_u$ and $OBF$. When adding a keyword is required, a DO runs the above algorithm and sends the result to the CCS to be stored in SBF and updates her $CBF_u$. On the other hand, to remove a keyword, the DO calculates a removal BF $(RBF_u)$ following the same procedures. Then, the DO compares $RBF_u$ with $CBF_u$ to see if any index in the $RBF_u$ has a value more than one in the corresponding index in $CBF_u$. For the ones for which $CBF_u$ has a value more than one, the user first flips the bit from 1 to 0 in $RBF_u$, then picks a random blinding element from $OBF$ where its corresponding value in $CBF_u$ is zero and updates $RBF_u$ with that to obtain $RBF'_u$. Note that the DO should also update $CBF_u$ using $RBF_u$ by decrementing the corresponding $r$ elements by 1. $OBF$ should also be updated if necessary. Finally, the DO sends the removal request along with $RBF'_u$ and $\xi ' \{\xi (MI_u)|| C(BF_u)|| PS_a\}$ to the server. 

Our scheme fully supports AND queries while it faces certain boundaries for OR queries. For queries on the AND of multiple keywords, a PSA runs the \textit{BuildIndex} algorithm on all the keywords and calculates $BF_u$. Instead of sending $L_{W_i}$, the PSA sends $\xi ' \{BF_u\}$ to the CCS/MCs. The \textit{SearchIndex} finds the intersection of indexes marked in the query and sends back the result. Due to space limitation, the full construction of multi-keyword search process for AND/OR queries will be presented in a separate work. 

\section{Security Analysis}
Our algorithm is semantically secure against a chosen keyword attack (IND-CKA): an attacker cannot learn anything about a set $W_u$ from its $BF_u$ for two main reasons. First, we use HMAC as our pseudo-random function; an adversary has negligible advantage to break this. Second, comparing two $BF_u$s, since both include $q$ elements by adding Obfuscating Elements (OE), the attacker will not learn which index contains more $W_j$ than  the other. 

To further quantify the influence of OEs, we are interested in the probability of an event $\varpi$ in which for any two users $a\text{ and }b$ with distinct sets of keywords $W_{u_a}\neq W_{u_b}$, after indexing $2q$ elements into an SBF, at least $r$ elements of user $a$ intersect with $r$ elements of user $b$ in the SBF. 
Higher $Pr(\varpi)$ makes more confusion for the server, thus providing more privacy. Following equation presents this probability $Pr(\varpi)=1-\sum\limits_{k=0}^{r-1}\dfrac{{\lambda \choose k}\times {m-\lambda \choose \lambda-k}}{{m \choose \lambda}}$,  

where $\lambda\simeq m-m\times e^{-rq/m}$ is the number of distinct elements after indexing $q$ items into an SBF. 
For example, 
if $l= 100$, $r=10$, $|\gamma|=1$, $q=15$, then, $m =1443$ and $\lambda =142$. Thus, $Pr(\varpi)\simeq 91.5$ percent. This means that even if $r$ elements of two distinct $BF_u$s intersect, with probability of 91.5 percent, those belong to two distinct sets of keywords. 
However, unnecessary overlapping of items might cause inaccurate results. Thus, we also need to calculate the probability of an event $\psi$ in which exactly $r$ elements in $BF_{u_a}$ intersect with those $r$ indexes in $BF_{u_b}$ where $W_ j\in W_{u_b}$ have mapped. Such probability is
$Pr(\psi)=\frac{q}{{\lambda \choose r}}\times \sum\limits_{k=r}^{\lambda}\dfrac{{\lambda \choose k}\times {m-\lambda \choose \lambda-k}\times {k \choose r}}{{m \choose \lambda}}$.
For the same parameters as the above example, $Pr(\psi)\simeq 1.6\times 10^{-10}$. This shows that our scheme is able to provide accurate search results with high probability and at the same time protects privacy of DOs through the use of OEs. 
Due to space constraints, formal proofs of IND-CKA security, $Pr(\varpi)$, and $Pr(\psi)$ are left for a separate paper.

In our scheme, a CCS or MC is not able to deduce which buffers in SBF are the target for a specific keyword. 
Our algorithm prevents dictionary attack, first, by randomizing the input value of the BF hash functions as a result of using HMAC; second, by utilizing the location parameter $\gamma$, in step 1-b of \textit{Buildindex} algorithm. The higher $\gamma$ causes more uniform distribution of records storage in the buffers of SBF.

Last but not least, confidentiality of $MI$ and search queries are provided via an encryption algorithm.
In addition, compared to the work in \cite{6683024}, our algorithm does not rely on a private server in order to build an index, generate search queries and provide privacy for them.

\section{Performance Analysis}

In this section, we will explore communication overhead, computation complexity, and memory usage. We simulated our scheme using the Java programming language on a desktop computer with the Ubuntu operating system. The PC is running on a core i3 CPU with a processing speed of 3.3 GHz. 

\subsection{Communication overhead}
The communication overhead from a DO to a CCS/MC is $ |\xi '\{\xi (MI_u)|| C(BF_u)|| PS_a\}|$. The maximum size of $|MI_u|$ is obtained when $q=|W_u|$. For every element in $W_u$ and the set $\{U_{P_{ID}}, CS_x, \sigma _u, PS_a\}$, we use 160 bits representation (e.g., using SHA1 as the generator). Also, we use Elliptic Curve Cryptography (ECC) 256 bits for $\xi()$ and AES 128 bits for $\xi '()$. Suppose, $q=15$, $|BF_u|=30$ kbits, and $r=10$, then $C(BF_u)$ can decrease $|BF_u|$ by approximately 92 percent \cite{goh2003secure}. Therefore, the maximum communication overhead will be less than 6 kbits. Note that the low communication overhead from the user to the server along with addition and removal capabilities of our system, allow our scheme to be compatible with dynamic scenarios where individuals constantly move from one location to another and need to update the SBF.

The communication overhead from a CCS/MC to a PSA is proportional to the number of files tagged with the queried keyword in a specific area (i.e., $|t_{W_i}|$). Thus, the overhead equals $|t_{W_i}|\times |\xi '\{\xi (MI_u)\}|$. However, $\gamma$ has a significant influence on $|t_{W_i}|$. Suppose, in a zone with pseudonym $PS_a$, there are $\gamma>1$ specified locations. Thus, two DOs with the same symptom like Asthma, but in different locations, would target different buffers in the SBF. Thus, searching for Asthma in one location, only retrieves one of the DOs. This produces a more accurate result and decreased communication overhead. 

\begin{figure}[t]
\centering
\includegraphics[width=0.8\linewidth,height=0.3\textheight]{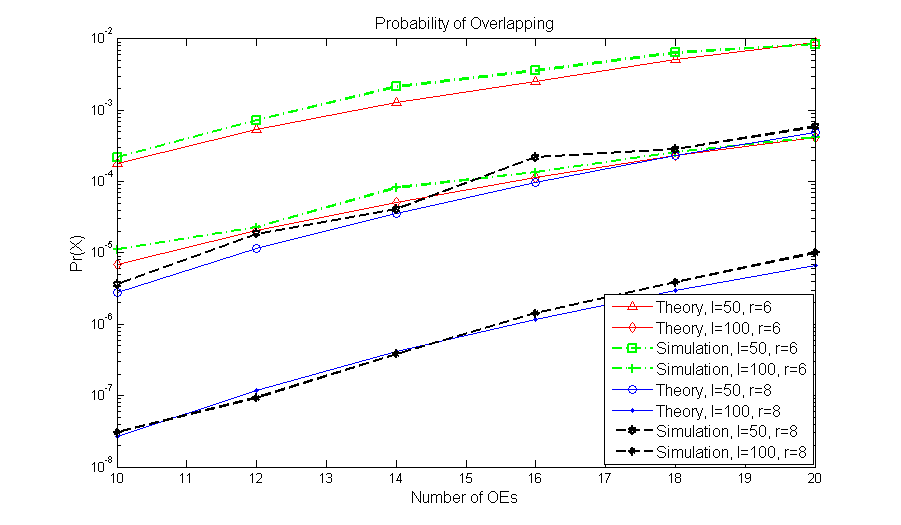}
\caption{Overlapping Probability}
\label{OP}
\end{figure}

However, a high number of OEs may affect accuracy of the results and increase communication overhead. Here, we investigate an event $\chi$ in which OEs may intersect with one or more of the $l$ keywords of the system in an SBF.
Note that the relationship among parameters of BF will be changed to $m=\frac{lr|\gamma|}{\ln 2}$ as a result of using $\gamma$. An upper-bound for the probability of the event $\chi$ occurs when DOs only insert OEs instead of $W_j$. Such probability is 
$P(\chi)\leq t\times {\lambda \choose r}\times l\times \gamma \times \frac{r!}{m^r}, $
where $\lambda$ is the same as above, and $t$ is the number of registered individuals. Fig. \ref{OP} shows the comparison between theoretical and simulation results for $P(\chi)$ when $t=1$ and $|\gamma|=1$.  

For each step in our simulation, we iterated 100 times and calculated the average value. In Fig. \ref{OP}, $r$ and $l$ take two different values. 
The figure illustrates that when the number of OEs increases, $Pr(\chi)$ increases. The maximum probability of approximately 0.8 percent occurs when $l=50$, $m=|SBF|=432$, $r=6$, and we added $|OE|=20$ to SBF. When $r$ is constant but $l$ and $|SBF|$ increase, $Pr(\chi)$ decreases. In addition, when $r$ increases, for the same $l$, the probability decreases. For $|OE|=15$, $l=100$, $|SBF|=1443$, $t=1000$, and $r=10$, theoretically $Pr(\chi)\leq 6.4\times 10^{-6}$. This result shows that the use of OEs in our scheme does not interfere with accuracy of search results and provides good levels of privacy. Note that $|\gamma|>1$ decreases $P(\chi)$ by a factor of $1/{\gamma ^{r-1}}$ since it causes $m$ to increase. 
\subsection{Memory}

The memory usage at CCS/MC equals $M=|SBF|\times\beta \times \tau$, where $|SBF|$ is the length of an SBF and $\beta$ is the maximum number of files inserted in one buffer, and $\tau =|\xi '(MI_u)|$ is the maximum size of each file in a buffer. Before we quantify $M$, we need to calculate the probability of an event in which one buffer overflows.

\begin{figure}[t]
\centering
\includegraphics[width=0.8\linewidth,height=0.5\textheight]{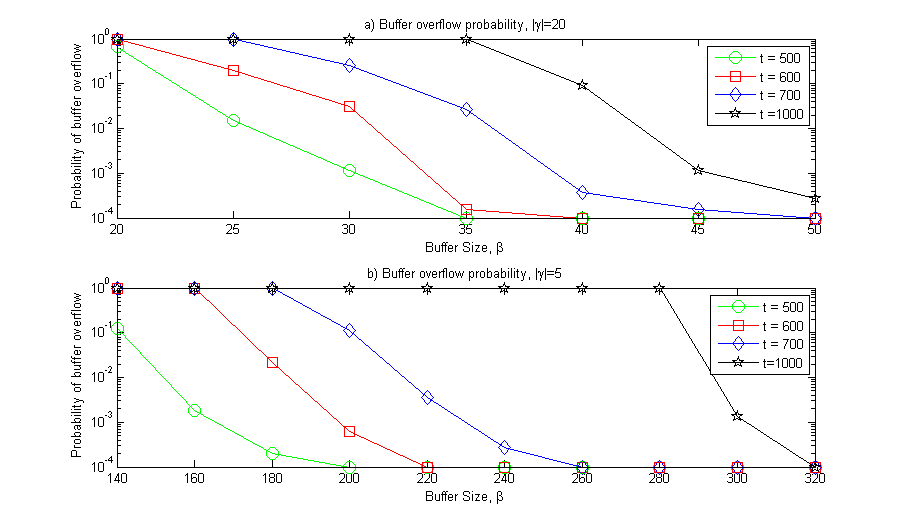}
\caption{Probability of buffer overflow, a) $|\gamma|=20$, b) $|\gamma|=5$}
\label{BOF5-20}
\end{figure}

Fig. \ref{BOF5-20} shows the buffer overflow probability. The figure depicts the buffer size requirements when the number of DOs increases from 500 to 1000 individuals and $|\gamma|$ changes from 5 to 20. Suppose, $l=100$, $r=10$, thus $|SBF|=\{28854, 7214\}$ for  $|\gamma| =\{20, 5\}$ respectively. Obviously, if buffer size increases, the probability of overflow decreases. Fig. \ref{BOF5-20}  $a$ shows that when $t=500$ and $\beta =20$, the overflow probability is approximately 67 percent. However, when $\beta$ increases to 35, the overflow probability drops to approximately $10^{-4}$. For $t=1000$, overflow probability is 1 until $\beta= 35$. But, increasing $\beta$ to 50 makes the overflow probability decrease to approximately $2.7\times 10^{-4}$. 

Fig. \ref{BOF5-20} $b$ depicts the situation where $|\gamma|=5$. 
For $t=1000$, using $\beta \geq 280$ the overflow probability decreases from 1 to approximately $10^{-4}$ for $\beta=320$. Comparing the two graphs in Fig. \ref{BOF5-20} shows that when $|\gamma|$ raises 4 times, the buffer size requirements falls approximately more than 5 times.  

Therefore, suppose $t=600$, $\tau =5Kbits$, $|\gamma|=20$, $l=100$, $r=10$, and  $\beta = 50$, $M\simeq 881 MB$. This result shows that our secure data storage structure is affordable even for MCs with resource constrained devices. 

\begin{table*}[t]
	\caption{Computation Complexity}
	\resizebox{\textwidth}{0.025\textheight}{
		\begin{tabular}{|c|c|c|c|c|c|}
			\hline Scheme & Buildindex ($t$ DOs)			  & Search (PSAs) & Search (CS) & Add (DO/CS) &Remove (DO/CS)\\ 
			\hline Ours & $t\times[(q\times 3r)H+\xi]$  & $rH +|t_{W_i}|\times D$ & $rH+I$ & $rH$ & $rH$ \\
			\hline \cite{6683024} & $t\times [(2q+2)PRP+q\times\xi]$ &    $2PRP$  & $I+|t_{W_i}|\times D$ & $2PRP + |t_{W_i}|\times D+|t_{W_i}+1|\times\xi$ & $2PRP + |t_{W_i}|\times D+|t_{W_i}-1|\times\xi$ \\
			\hline
			
		\end{tabular} 
		}

	\label{computation}
	
\end{table*}

\subsection{Computational overhead and delay}

Table \ref{computation} shows the computational complexity for each procedure at the user side and server side in comparison with the work in \cite{6683024}. In terms of computational complexity, we used HMAC in building the index and search procedure which was done in a distributed way with very low delay. The user in \cite{6683024} is a private server to which the computations of $t$ registered individuals are outsourced.  $H$ is used to show HMAC computation and $PRP$ stands for a pseudorandom permutation function such as AES. $t_{W_i}$ is the number of users who have the same keyword. $\xi, D$ are used to show encryption and decryption processes respectively. Finally, $I$ is used to show the intersection operation between $r$ buffers.

Increasing $\gamma$ decreases $\beta$ requirements which implies that the number of intersection operations and search results will also be decreased. Consequently, computational complexity decreases. 
Furthermore, our scheme imposes very low computation burden for addition and deletion processes in comparison with the work in \cite{6683024} in which the public server needs to decrypt the entire linked list and then modify it for any single alteration. This indicates the applicability of our scheme for dynamic situations where the cost of updating needs to be limited.

In general, the data retrieval process consists of two procedures, search over encrypted data and decryption. In our scheme and \cite{6683024}, search takes place merely over the number of files containing the keyword and not the entire database (i.e., $O(1)$ delay) which is far better that the works in \cite{li2011authorized, liu2012secure}. Note that $O(1)$ delay has significant impact on data access under critical circumstances where the size of a database is large or an immediate response is required. Our scheme decrypts ECC and AES ciphertext messages, whereas the methods in \cite{liu2012secure, dong2011shared, barua2011peace, zhou2014psmpa, li2013scalable, lu2013spoc, liang2011pec, liang2012healthshare, zhang2014phda} involve PBC which requires more computational resources. 

\section{Conclusion}
In PS situations, privacy preservation, context, and location-aware information are required. 
Existing works did not address such requirements in PSNs.
In this work, we proposed a storage bloom filter and modified a secure index algorithm to provide data availability with regards to PSN requirements. Our search process imposes $O(1)$ delay which is ideal for PS situations. In addition, communication complexity is very low from a DO to a CS and it is proportional to the number of files containing the search query in reverse direction. The memory usage is also affordable even for MCs with limited resources. We used a location parameter $\gamma$ with which we decreased the buffer size and the number of search outcomes. The latter decreases communication and computational complexities and delay. To the best of our knowledge, this work is the first to address such features in PSNs. 

\bibliographystyle{IEEEtran}
\bibliography{cos-arxiv-version}

\end{document}